\documentclass[useAMS,usenatbib]{mn2e}

\usepackage{epsfig}
\usepackage{longtable}
\usepackage{times}
\def \inte {{$INTEGRAL$}}

\def \sw {{\em Swift}}
\def \src {\mbox{IGR~J08408--4503}}

\def \hcm {\hbox {\ifmmode $ atom cm$^{-2}\else atom cm$^{-2}$\fi}}

\def \ATel {Astron.\ Tel.}
\def \apj {ApJ}
\def \apjl {ApJL}
\def \aap {A\&A}

\def \pasj {PASJ}
\def \gcn {GCN Circ.}

\title[Flaring activity in IGR J08408-4503]{Multiple flaring activity in 
the supergiant fast X-ray transient \src\ observed with {\it Swift} }
\author[P. Romano et al.]{P.\ Romano$^{1}$, L.\ Sidoli$^{2}$, G.\ Cusumano$^{1}$, P.A.\ Evans$^{3}$,  L.\ Ducci$^{4,2}$, H.A.\ Krimm$^{5,6}$,
\newauthor   S.\ Vercellone$^{2}$, K.L.\ Page$^{3}$,
	A.P.\ Beardmore$^{3}$, D.N.~Burrows$^{7}$, J.A.~Kennea$^{7}$, N.~Gehrels$^{6}$, 
\newauthor V.\ La Parola$^{1}$, V.\ Mangano$^{1}$ \\
$^{1}$INAF, Istituto di Astrofisica Spaziale e Fisica Cosmica,
        Via U.\ La Malfa 153, I-90146 Palermo, Italy\\
$^{2}$INAF, Istituto di Astrofisica Spaziale e Fisica Cosmica,
	Via E.\ Bassini 15,   I-20133 Milano,  Italy\\
$^{3}$Department of Physics \& Astronomy, University of Leicester, LE1 7RH, UK\\
$^{4}$Dipartimento di Fisica e Matematica, Universit\`a dell'Insubria, Via
Valleggio 11, I-22100 Como, Italy \\
$^{5}$Universities Space Research Association, Columbia, MD, USA \\
$^{6}$NASA/Goddard Space Flight Center, Greenbelt, MD 20771, USA\\
$^{7}$Department of Astronomy and Astrophysics, Pennsylvania State 
             University, University Park, PA 16802, USA\\
}

\begin{document}

\date{}

\pagerange{\pageref{firstpage}--\pageref{lastpage}} \pubyear{2008}

\maketitle

\label{firstpage}

\begin{abstract}
\src\ is a supergiant fast X--ray transient discovered in 2006 
with a confirmed association  with a O8.5Ib(f) supergiant star, HD~74194.
We report on the  analysis of two outbursts caught by {\it Swift}/BAT 
on 2006 October 4 and 2008 July 5, 
and followed up at softer energies with {\it Swift}/XRT. 
The 2008 XRT light curve shows a multiple-peaked structure with an initial bright 
flare that reached a flux of $\sim 10^{-9}$ erg\,cm$^{-2}$\,s$^{-1}$ (2--10\,keV),
followed by two equally bright flares within 75\,ks. 
The spectral characteristics 
of the flares differ dramatically, with most of the difference, as derived via 
time-resolved spectroscopy,  being due to absorbing column variations. 
We observe a gradual decrease of the 
$N_{\rm H}$, derived with a fit using absorbed power law model, as time passes. 
We interpret these $N_{\rm H}$ variations as due to an ionization effect 
produced by the first flare, resulting in a significant decrease in the 
measured column density towards the source. 
The durations of the flares, as well as the times of the outbursts suggest 
that the orbital period is $\sim$35~days, if the flaring activity is interpreted 
within the framework of the \citet{Sidoli2007} model
with the outbursts triggered by the neutron star passage inside an 
equatorial wind inclined with respect to the orbital plane.
\end{abstract}


	\section{Introduction\label{igr08408:intro}}

\src\ belongs to the class of the supergiant fast X--ray transients (SFXTs), 
which are transient sources in 
high mass X--ray binaries (HMXBs) associated with blue supergiant companions 
(see e.g., \citealt{Smith2006aa}). 
Several members of this class have been discovered 
since the launch in 2002 of the \inte\
satellite, thanks to its  monitoring of the Galactic plane  performed with a 
large field of view and a 
good sensitivity at hard X-rays. 
Indeed, several of these transients are highly 
absorbed in X--rays, and previous missions
had failed to detect them \citep[see e.g.][]{Sguera2005}.
Two members of this class are X--ray pulsars: IGR~J11215--5952 \citep{Swank2007} 
and AX~J1841.0$-$0536/IGR~J18410$-$0535 \citep{Bamba2001}.
The X--ray spectra of SFXTs are very similar to those typically displayed 
by accreting X--ray pulsars, with
a flat power law spectrum below 10 keV and a high energy cut-off around 10--30~keV 
(see \citealt{Romano2008:sfxts_paperII}, for
the first truly simultaneous wide-band X--ray spectrum of a SFXT, from 0.3 to 100 keV). 
This spectral similarity suggests that the other members of 
the new class of HMXBs probably host a neutron star, as well.

The emerging picture of the SFXTs as a class, especially thanks to 
more sensitive monitoring
performed with \sw\ of four SFXTs [see \citet{Sidoli2008:sfxts_paperI} 
for the strategy of this on-going campaign], 
is that they  sporadically undergo bright flares, up to
peak luminosities of 10$^{36}$--10$^{37}$~erg~s$^{-1}$, with a duration of a few
hours for each single flare.
This bright flaring activity is actually part of a much longer duration outburst event, 
 lasting days [see \citet{Romano2007} and \citet{Sidoli2008:sfxts_paperIII}].
Their long-term emission is characterized by 
a much lower intensity and flaring activity, with an average luminosity 
of 10$^{33}$--10$^{34}$~erg~s$^{-1}$
\citep{Sidoli2008:sfxts_paperI}. 
The true quiescence has been rarely observed, 
and is characterized
by a very soft spectrum, with a luminosity as low as 10$^{32}$~erg~s$^{-1}$ 
(\citealt{zand2005}, \citealt{Leyder2007}).

\src\ was discovered on 2006 May 15, during a short bright 
flare with a duration of 900~s, reaching a peak flux of 250 mCrab in the 
20--40 keV energy band \citep{Gotz2006:08408-4503discovery}. 
Analysis of archival \inte\ observations of the source field led to the discovery
that \src\ is a recurrent transient, being also active on 2003 July 1 \citep{Mereghetti:08408-4503}.
The refined position with \sw/XRT \citep{Kennea2006:08408-4503},
during a 
Target of Opportunity observation (ToO)
performed about one week after the discovery flare,
allowed the confirmation of the association  with a O8.5Ib(f) supergiant star, HD~74194, 
located in the Vela region \citep{Masetti2006:08408-4503} at a distance of about 3\,kpc.
Optical spectroscopy of HD~74194 just a few days after the 2006 May 15 flare 
revealed variability in the H${\alpha}$ profile and a radial velocity variation 
in the He\,I and He\,II absorption lines with an amplitude of about  35 km\,s$^{-1}$.
The study of three additional flares observed with \inte\ and \sw\ by \citet{Gotz2007:08408-4503} led
these authors to suggest that the orbital period of \src\ is probably of the order of 1 yr, the spin period 
of the putative neutron star could be of order
of hours, and the surface magnetic field is probably around 10$^{13}$~G.

In this paper we report on the detailed data analysis of both the 
2006 October 4 and the 2008 July 5 outbursts caught by \sw/BAT 
and followed up at softer energies with \sw/XRT.

 	 \section{Observations and Data Reduction\label{igr08408:dataredu}}

\src\  triggered the {\it Swift}/BAT on 2006 October 10 14:45:43 UT 
(image trigger=232309) 
and on 2008 July 5 21:14:15 UT 
\citep[image trigger=316063,][]{Ward2008:08408-4503}, 
respectively.
On both occasions, {\it Swift} slewed to the
target, allowing the narrow field instruments (NFIs) to be
pointing at the target $\sim1885$ s and $\sim130$\,s after the 
trigger, respectively. 
We note that  this source is normally not detectable in the 
BAT Transient Monitor, but has shown a few untriggered outbursts over the past 
three years. 
Table~\ref{igr08408:tab:alldata} 
reports the log of the \sw/XRT observations 
used for this work.

\begin{table*}
 \begin{center}
 \caption{Observation log.\label{igr08408:tab:alldata} }
 \begin{tabular}{llllr}
 \hline
 \hline
 \noalign{\smallskip}
Sequence & Instrument & Start time (UT) & End time (UT) &  Exposure$^{\mathrm{a}}$ \\
  & /Mode & (yyyy-mm-dd hh:mm:ss) & (yyyy-mm-dd hh:mm:ss) & (s) \\
  \noalign{\smallskip}
 \hline
 \noalign{\smallskip}
00232309000	&XRT/WT &	2006-10-04 15:17:16	&	2006-10-04 16:51:12	&	180	\\
00232309000	&XRT/PC &	2006-10-04 15:20:08	&	2006-10-04 17:05:56	&	1472	\\
00232309001	&XRT/PC &	2006-10-04 18:27:05	&	2006-10-04 18:42:58	&	951	\\
00316063000	&XRT/PC &	2008-07-05 21:18:23	&	2008-07-06 18:29:51	&	29956	\\
00316063000	&XRT/WT &	2008-07-05 21:16:31	&	2008-07-06 17:55:16	&	447	\\
00030707003	&XRT/PC &	2008-07-08 13:26:02	&	2008-07-08 18:23:18	&	1978	\\
00030707004	&XRT/PC &	2008-07-09 18:26:26	&	2008-07-09 18:43:04	&	998	\\
00030707005	&XRT/PC &	2008-07-10 23:12:51	&	2008-07-10 23:28:56	&	965	\\
00030707006	&XRT/PC &	2008-07-11 00:55:02	&	2008-07-11 02:41:57	&	1308	\\
00030707007	&XRT/PC	&	2008-07-12 20:25:43	&	2008-07-12 22:14:56	&	1319	\\
00030707008	&XRT/PC	&	2008-07-14 20:28:10	&	2008-07-14 23:47:51	&	1203	\\
00030707009	&XRT/PC	&	2008-07-15 04:30:10	&	2008-07-15 23:51:55	&	2384	\\
00030707010	&XRT/PC &	2008-07-16 19:03:10	&	2008-07-16 23:59:57	&	1474	\\
00030707011	&XRT/PC &	2008-07-17 22:27:20	&	2008-07-17 22:38:25	&	664	\\
00030707012     &XRT/PC &       2008-07-20 11:37:03     &       2008-07-20 11:56:56     &       1193  \\  
  \noalign{\smallskip}
  \hline
  \end{tabular}
  \end{center}
  \begin{list}{}{}
  \item[$^{\mathrm{a}}$]{The exposure time is spread over several snapshots  
	(single continuous pointings at the target) during each observation.}
  \end{list}  
\end{table*}

\begin{figure}
\begin{center}
\centerline{\includegraphics[width=9cm,angle=0]{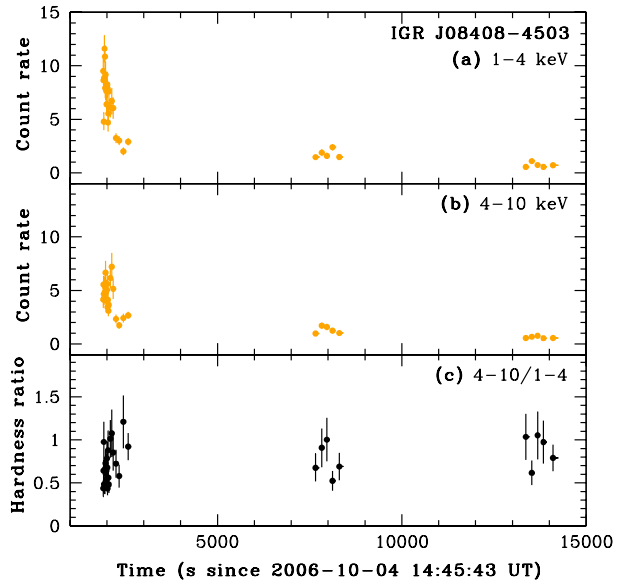}}
\caption{XRT light curves of the 2006 October 4 outburst
		background-subtracted and corrected for pile-up, PSF losses, and vignetting,
	rebinned with a minimum of 30 counts per bin:
		{\bf a)} 1--4\,keV band; 
		{\bf b)}  4--10\,keV energy band. 
		{\bf c)}  4--10/1--4 hardness ratio.
	}
\label{igr08408:fig:2006lcv}
\end{center}
\end{figure}

\begin{figure}
\begin{center}
\centerline{\includegraphics[width=9cm,angle=0]{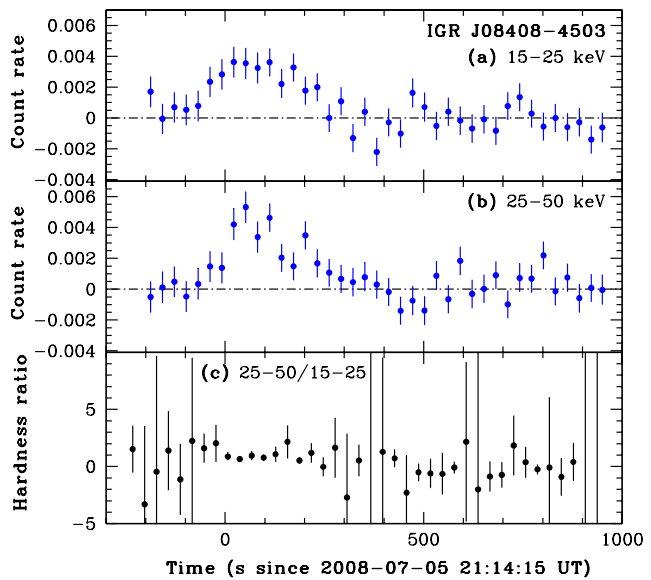}}
\caption{BAT light curves of the 2008 July 5 outburst,
rebinned at a $S/N=10$.  
 		{\bf a)} 15--25\,keV band (units of counts s$^{-1}$ det$^{-1}$);  
		{\bf b)} 25--50\,keV band (counts s$^{-1}$ det$^{-1}$); 
		 {\bf c)} 25--50/15--25 hardness ratio.
	}
\label{igr08408:fig:2008bat}
\end{center}
\end{figure}

The XRT data were processed with standard procedures 
({\sc xrtpipeline} v0.11.6), filtering and screening criteria by using 
{\sc FTOOLS} in the {\sc Heasoft} package (v.6.4).  
We considered both WT and PC data, 
and selected event grades 0--2 and 0--12, respectively 
(\citealt{Burrows2005:XRTmnras}).
When appropriate (observations 00232309000 and 00316063000), 
we corrected for pile-up 
by determining the size of the PSF core affected 
by comparing the observed and nominal PSF \citep{vaughan2006:050315},
and excluding from the analysis all the events that fell within that
region. 
Ancillary response files were generated with {\sc xrtmkarf},
and they account for different extraction regions, vignetting, and
PSF corrections. We used the spectral redistribution matrices
v010 in CALDB. 

The BAT always observed \src\ simultaneously with XRT, so 
survey data products, in the form of Detector Plane
Histograms (DPH) with typical integration time of 
$\sim 300$\,s, are available. Furthermore, event data 
were collected during observation 00316063000. 
The BAT data were analysed using the standard BAT analysis 
software distributed within {\sc FTOOLS}.
BAT mask-weighted spectra were extracted over the time
intervals simultaneous with XRT data (see Sect.~\ref{igr08408:spectra}).
Response matrices were generated with {\sc batdrmgen} 
using the latest spectral redistribution matrices and the default computation
method ({\sc method=mean}) accessing calibration parameters
\citep{BATmanual2007} that are averaged over the entire
BAT field of view (FOV). 
For our spectral fitting (XSPEC v11.3.2)
we applied an energy-dependent
systematic error, amounting to 4--5\%
through all the 15$-$90\,keV energy band \citep{BATmanual2007}.
Inspection of the BAT FOV shows that no source substantially
brighter than \src\ was visibile during both the 2006 and 2008 observations,
and we can estimate that possible contamination from other
sources in the FOV cannot exceed 5\,\%.

UVOT data were also collected simultaneously with the other instruments;
however, due to the brightness of HD~74194 (LM Vel), the data were not usable 
in order to study the variability of the optical counterpart.

All quoted uncertainties are given at 90\% confidence level for 
one interesting parameter unless otherwise stated. 
The spectral indices are parameterized as  
$F_{\nu} \propto \nu^{-\alpha}$, 
where $F_{\nu}$ (erg cm$^{-2}$ s$^{-1}$ Hz$^{-1}$) is the 
flux density as a function of frequency $\nu$; 
we adopt $\Gamma = \alpha +1$ as the photon index, 
$N(E) \propto E^{-\Gamma}$ (ph cm$^{-2}$ s$^{-1}$ keV$^{-1}$). 
Times in the light curves and the text are referred to their respective 
triggers.

  	\section{Analysis and Results\label{igr08408:dataanal}}

\begin{figure}
\begin{center}
\centerline{\includegraphics[width=9cm,angle=0]{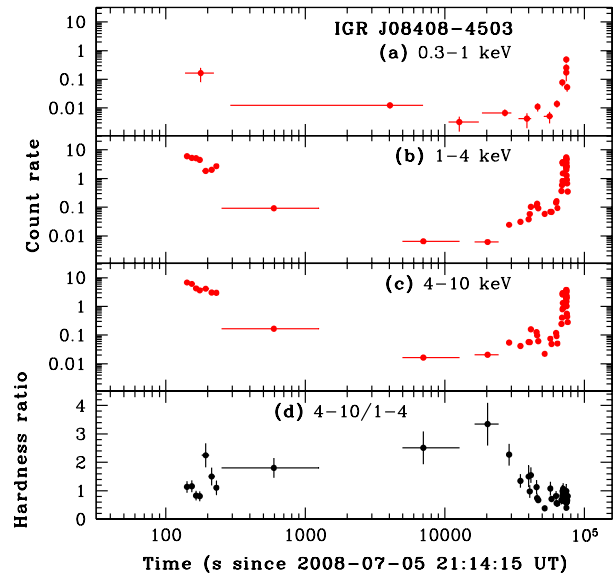}}
\caption{XRT light curves of the 2008 July 5 outburst 
	rebinned with a minimum of 30 counts per bin:
		{\bf a)} 0.3--1\,keV band. 
		{\bf b)} in the 1--4\,keV band; 
		{\bf c)} in the 4--10\,keV energy band; 
		{\bf d)} 4--10/1--4 hardness ratio. 
	}
\label{igr08408:fig:2008lcv}
\end{center}
\end{figure}

  	\subsection{Light curves\label{igr08408:lcvs}}

For both observations 00232309000 (2006 outburst) and 00316063000 (2008), 
XRT light curves were extracted in several energy bands 
using the procedures described in \citet{Evans2007:repository}. 
Mask-tagged BAT light curves were created in the standard 4 energy bands, 
15--25, 25--50, 50--100, 100--150 keV, for observation 00316063000, 
and rebinned to achieve a signal-to-noise ratio (S/N) of 10. 
Fig.~\ref{igr08408:fig:2006lcv} shows the XRT light curves 
for the 2006 outburst, while Figs.~\ref{igr08408:fig:2008bat} 
and \ref{igr08408:fig:2008lcv} show the BAT and XRT light curves for the 
2008 outburst, respectively.
As shown in Fig.~\ref{igr08408:fig:2006lcv}, the late start of the 
2006 XRT observation (T+1885\,s) only allowed us to follow the descent of the outburst.
In contrast, the XRT 2008 data offer a much better coverage and show
a remarkable second and third flares, which reached the same intensity as the first one.

Figs.~\ref{igr08408:fig:2006lcv}c and \ref{igr08408:fig:2008lcv}d show the 
4--10/1--4 hardness ratios. Fitting the 2006 XRT hardness ratio against a 
constant yields a value of $0.60\pm0.02$ 
($\chi^2_{\nu}=1.64$ for 33 degrees of freedom, 
d.o.f.), which corresponds a $\sim 2.5\sigma$ variation.
In the case of the 2008 XRT data we see a definite trend for a softening of the 
spectrum as the source becomes brighter, which is reminiscent 
of what was observed, albeit with lower statistics, during the 2008 March 19 outburst of
IGR~J16479$-$4514 \citep{Romano2008:sfxts_paperII}. 
No significant variation in the 2008 BAT hardness ratio is evident 
(Fig.~\ref{igr08408:fig:2008bat}c). 
Fitting the hardness ratio as a function of time to a constant model
we obtain a value of $0.58\pm0.11$ ($\chi^2_{\nu}=0.56$ for 39 d.o.f.).  

A timing analysis was performed on WT data,  
after having converted the event arrival times to the Solar System 
Barycentric frame. We searched for coherent periodicity, 
but found no evidence for pulsations in the range 10\,ms--100\,s.

  	\subsection{Broad-band spectra\label{igr08408:spectra}}

\begin{figure}
\begin{center}
\centerline{\includegraphics[width=6cm,angle=270]{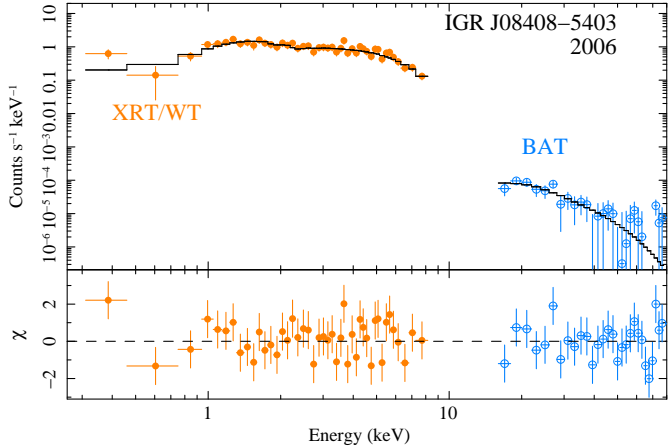}}
\caption{Spectroscopy of the 2006 October 4 outburst. 
		{\bf Top:} simultaneous fit of BAT and XRT/WT data
		with an absorbed power law with a high energy cutoff. 
		{\bf Bottom:} the residuals of the fit (in units of standard deviations). 
	}
\label{igr08408:fig:2006spec}
\end{center}
\end{figure}

\begin{figure}
\begin{center}
\centerline{\includegraphics[width=6cm,angle=270]{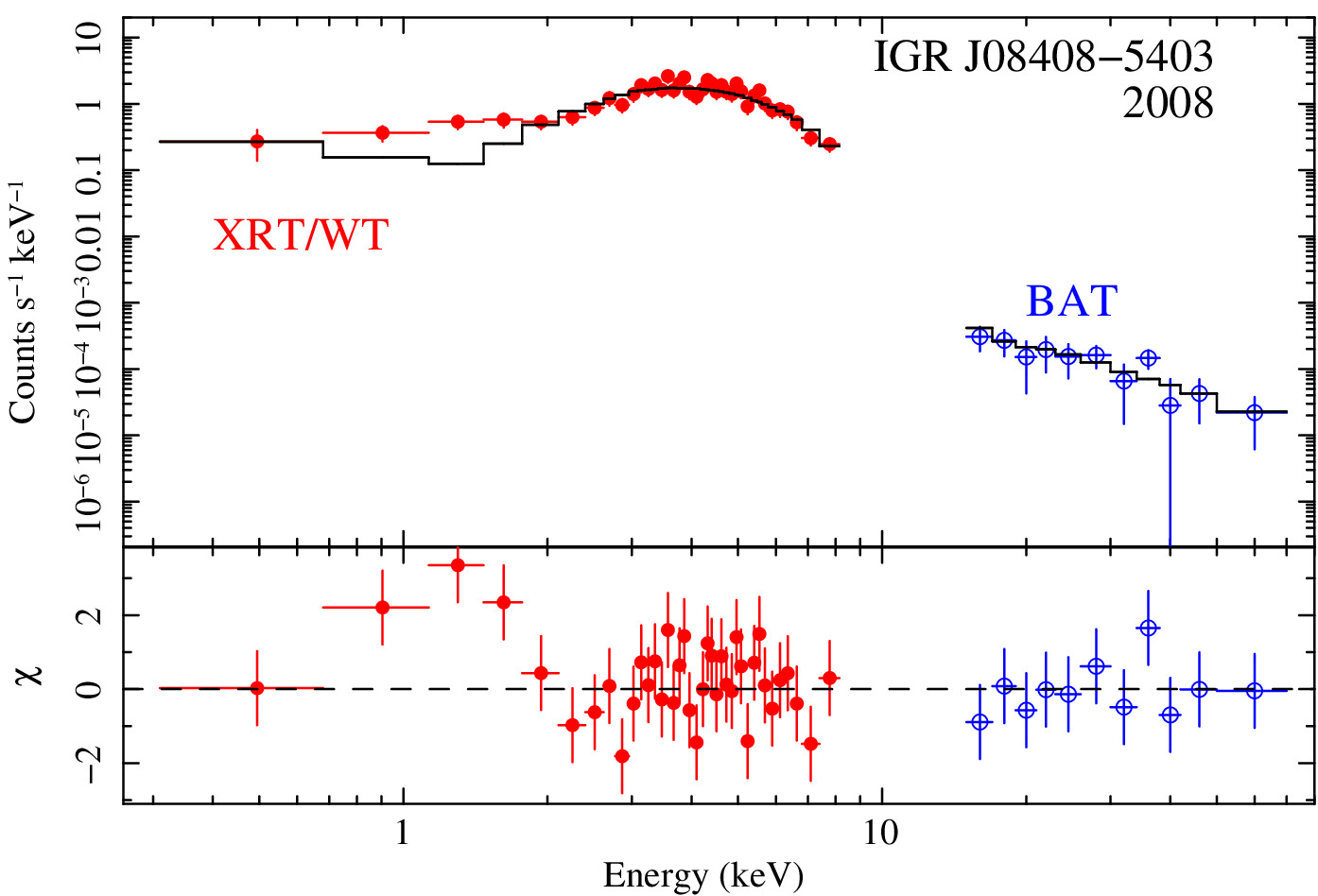}}
\caption{Spectroscopy of the 2008 July 5 outburst. 
		{\bf Top:} simultaneous fit of BAT and XRT/WT data
		with an absorbed power law with a high energy cutoff. 
		{\bf Bottom:} the residuals of the fit (in units of standard deviations).  
	}
\label{igr08408:fig:2008spec}
\end{center}
\end{figure}

For the 2006 October 4 outburst we extracted one BAT spectrum nearly
simultaneous with the XRT/WT data set and performed a broad-band fit to an 
absorbed power law with an exponential cutoff. 
The XRT data were rebinned with a minimum of 20 counts per energy 
bin to allow $\chi^2$ fitting, while the BAT data were rebinned to a $S/N>5$. 
We fit in the 0.3--10\,keV and 15--80\,keV energy ranges for the XRT and BAT data, 
respectively, with XSPEC (v11.3.2). 
Similarly, we extracted a BAT spectrum simultaneous with the XRT/WT data 
(137--246\,s) for the 2008 July 5 outburst.  
The best fit parameters are reported in Table~\ref{igr08408:tab:broadspecfits}, 
while the spectra are shown in Figs.~\ref{igr08408:fig:2006spec} and  
~\ref{igr08408:fig:2008spec}, respectively. 
We note that there are no BAT event data during the second and third flares of the 2008
light curve, since the source was already identified on-board as the current Automated Target 
and it did not reach the required count rate (twice the previous triggering rate 
plus 7 $\sigma$) to trigger the BAT again. 
We also note that the presence of a cutoff is required by the 
data, since the trend in the residuals of a simple power-law model 
clearly indicates the presence of curvature in the spectrum.

 \begin{table}
 \begin{center}
 \caption{Broad-band spectral fits of the BAT+XRT data.\label{igr08408:tab:broadspecfits} }
 \begin{tabular}{llrrr}
 \hline
 \hline
 \noalign{\smallskip}
  Year&$N_{\rm H}$ 		&  $\Gamma$ 	&  $E_{\rm c}$ &  $\chi^{2}_{\nu}$/d.o.f.\\
  &($10^{22}$ cm$^{-2}$)  & 		&  (keV) 	& \\
  \noalign{\smallskip}
 \hline
 \noalign{\smallskip}
   2006 	&$0.29_{-0.14}^{+0.17}$ & $0.31_{-0.23}^{+0.23}$ & $11_{-3}^{+4}$   & $0.903/69$  \\
   2008 	&$6.7_{-1.8}^{+2.0}$ & $1.36_{-0.56}^{+0.49}$ & $>14$  &   $1.191/45$ \\
  \noalign{\smallskip}
  \hline
  \end{tabular}
  \end{center}
  \end{table}

  	\subsection{Time-resolved spectra\label{igr08408:timespectra}}

\begin{figure}
\begin{center}
\centerline{\includegraphics[width=9.5cm,angle=0]{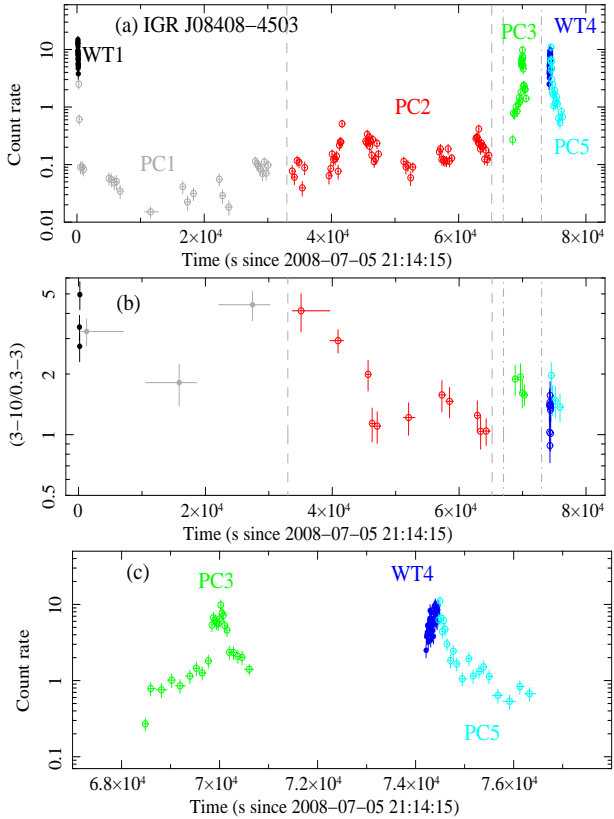}}
\vspace{-1.5truecm}
\caption{Time selections for XRT spectroscopy of the 2008 July 5 outburst. 
		{\bf (a):} XRT 0.3--10\,keV light curve; the intervals over which we performed spectroscopy
		are marked with vertical lines and labels. Filled
                circles are WT data, empty circles PC data.
		{\bf (b):} 3--10/0.3--3 keV hardness ratio. 
		{\bf (c):} detail of (a) to show the second and third flare; 
	}
\label{igr08408:fig:2008xrttimes}
\end{center}
\end{figure}

\begin{figure}
\begin{center}
\centerline{\includegraphics[width=9.5cm,angle=0]{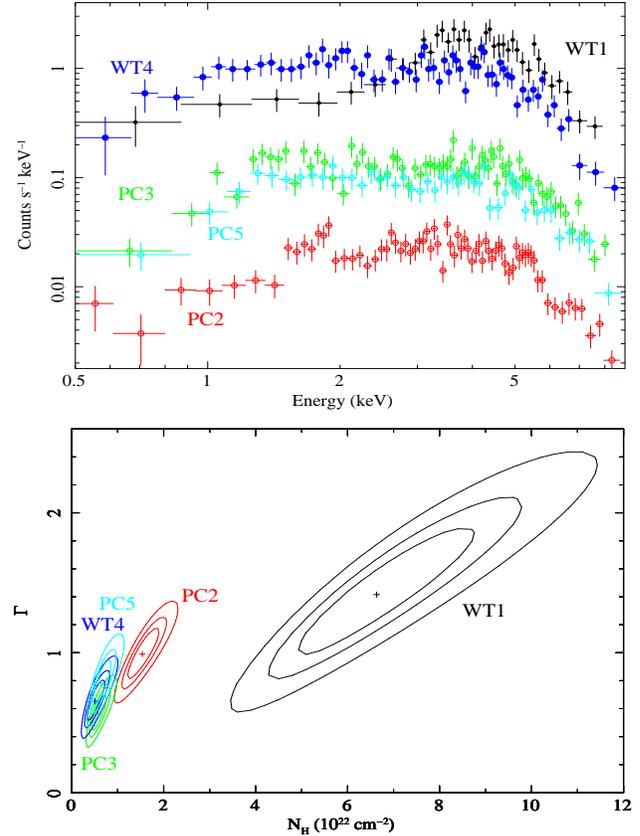}}
\vspace{-1.5truecm}
\caption{XRT time-selected spectroscopy of the 2008 July 5 outburst: 
		{\bf (top):} Spectra extracted in the different intervals.  
                {\bf (bottom):} 
	68\%, 90\% and 99\% confidence level contours adopting an absorbed power-law model.
	The labels refer to the time intervals shown in Fig.~\ref{igr08408:fig:2008xrttimes}
	and Table~\ref{igr08408:tab:timespecfits}. 
	}

\label{igr08408:fig:2008xrtcontours}
\end{center}
\end{figure}

Upon examination of the light curve and the available counting statistics, 
we selected different time bins over which we accumulated  spectra. 
For the 2006 data, we extracted one spectrum in WT mode and two in PC mode
(observations 00232309000 and 00232309001 in Table~\ref{igr08408:tab:alldata}). 
For the 2008 data, our time selections include (see Fig.~\ref{igr08408:fig:2008xrttimes}a,c, 
and Table~\ref{igr08408:tab:timespecfits}):  
{\it i)} the initial WT data (137--246\,s, hereafter WT1); 
{\it ii)} a low phase, with a mean CR $\sim3.5\times10^{-2}$ counts s$^{-1}$ 
            (4762--30180\,s, orbits 2--6 of observation 00316063000, PC1); 
{\it iii)} an intermediate phase, with a mean CR $\sim 0.1$ counts s$^{-1}$ 
         (33697--64903\,s, orbits 7--12,  PC2); 
{\it iv)} the second flare, (68429--70690\,s, orbit 13,  PC3);
{\it v)} the rise of the third flare (74208--74462\,s, orbit 14, WT4);
{\it vi)} the decay of the third flare (74463--76477\,s, orbit 14, PC5). 
The spectra were rebinned with a minimum of 
20 counts per energy bin to allow $\chi^2$ fitting (see Fig.~\ref{igr08408:fig:2008xrttimes}c). 
However,  the Cash statistic \citep{Cash1979} and spectrally 
unbinned data were used, due to the low counting statistics, to fit the PC1 spectrum 
 (the unbinned spectrum is not shown 
Fig.~\ref{igr08408:fig:2008xrtcontours}).
Each spectrum was fit in the 0.3--10\,keV energy range,
adopting an absorbed power law model. 
The best fit parameters are reported in Table~\ref{igr08408:tab:timespecfits}. 
Remarkably, we obtain significantly different fitting parameters for the 
data of the three flares, despite quite similar flux levels. 
Indeed, for the first one (WT1) we obtain a photon index 
$\Gamma_{\rm WT1}=1.42_{-0.47}^{+0.52}$ and an absorbing column density of 
$N_{\rm H \, WT1}=(6.63_{-1.88}^{+2.34})\times 10^{22}$ cm$^{-2}$, 
while the second and third ones (PC3, WT4/PC5) yielded 
$\Gamma_{\rm WT4}=0.65_{-0.15}^{+0.17}$ and 
$\Gamma_{\rm PC5}=0.78_{-0.17}^{+0.18}$, 
 and an absorbing column density of 
$N_{\rm H \,WT4}=(0.50_{-0.18}^{+0.24})\times 10^{22}$ cm$^{-2}$   
$N_{\rm H \,PC5}=(0.64_{-0.19}^{+0.24})\times 10^{22}$ cm$^{-2}$.
Fig.~\ref{igr08408:fig:2008xrtcontours} (bottom) shows the 
contour levels for the column density vs.\ the photon index, which we 
created to investigate these spectral variations.
As can be seen, the photon index range is restricted to $\Gamma \approx 1\pm0.5$,
while the measured absorbing column varies by an order of magnitude 
with a definite trend for a
decreasing $N_{\rm H}$ as time goes by. It must also be noted that the 2006 spectra,
taken about 2000\,s from the trigger, are quite similar to 
these `late' (PC3 onwards) spectra. 
We note the presence of a bump in the residuals at soft energies, around 
1\,keV (see Fig.~\ref{igr08408:fig:2008spec}).
This feature, which has a variable contribution to the spectra
and is stronger in WT1, is probably due to a combination of a slight spectral
variability throughout the time interval during which the spectrum was accumulated (see
Fig.~\ref{igr08408:fig:2008lcv}d and ~\ref{igr08408:fig:2008xrttimes}b), 
and some residual calibration uncertainties below 2\,keV in WT data 
\citep{Godet2008:xrtcalib}.
We are however confident that the latter do not 
weaken our findings of a decreasing $N_{\rm H}$, since their effect is
in the opposite direction and we shall not discuss this feature
any further.

	\section{Discussion and Conclusions\label{igr08408:discussion}}

	\subsection{Multiple flaring behaviour}

In this paper we report on two outbursts of the 
supergiant fast X-ray transient  \src\ observed by \sw\ on 
2006 October 4 and 2008 July 5. 
During the 2008 outburst, a multiple-flare behaviour 
is observed so clearly in a SFXT,  and for the first time with a 
large dynamic range, thanks to the good sampling of the source light curve.
The 2008 outburst emission is composed of three main bright flares, 
each exceeding 10$^{36}$~erg~s$^{-1}$, with a duration of a few hundred seconds. 
The flares are separated by $\sim$70~ks (between the first and second flare) 
and $\sim$4~ks (second to third flare).

 \begin{table*}
 \begin{center}
 \caption{Spectral fits of the XRT data.}
 \label{igr08408:tab:timespecfits}
 \begin{tabular}{llrrrrr}
 \hline
 \hline
 \noalign{\smallskip}
  Year &Spectrum &Times	&$N_{\rm H}$ 	& $\Gamma$ &Flux$^{\mathrm{a}}$    & $\chi^{2}_{\nu}$/d.o.f. \\
      &   &  (s) &($10^{22}$ cm$^{-2}$)  &   	&   & \\
  \noalign{\smallskip}
 \hline
 \noalign{\smallskip}
   2006&  WT	     & 1893-2064	& $0.37_{-0.14}^{+0.17}$  & $0.67_{-0.17}^{+0.17}$ &5.7   & $0.902/42$\\
       &  PC000      & 2064--8413	& $0.90_{-0.28}^{+0.35}$  & $0.88_{-0.22}^{+0.24}$ &1.4   & $1.477/31$\\
       &  PC001      & 13283--14234     & $0.47_{-0.27}^{+0.41}$  & $0.49_{-0.26}^{+0.30}$ &1.5   & $0.554/19$\\
       &  all        & 1893-14234	& $0.57_{-0.13}^{+0.15}$  & $0.69_{-0.12}^{+0.13}$ &      & $1.118/96$\\
   2008&  WT1 & 137--246	   & $6.63_{-1.88}^{+2.34}$  & $1.42_{-0.47}^{+0.52}$	  &13.4   &  $1.255/35$ \\
       &  PC1 & 4762--30180         &$11.08_{-4.20}^{+5.12}$ &$0.65_{-0.69}^{+0.75}$ &0.086 & 578.4 (55.78\%)$^{\mathrm{b}}$  \\
       &  PC2 & 33697--64903	   & $1.54_{-0.33}^{+0.40}$  & $0.99_{-0.19}^{+0.20}$	  &0.14   &  $1.423/70$\\
       &  PC3 & 68429--70690	   & $0.57_{-0.16}^{+0.20}$  & $0.57_{-0.14}^{+0.14}$	  &1.8    &  $1.368/70$	\\	   
       &  WT4 & 74208--74462	   & $0.50_{-0.18}^{+0.24}$  & $0.65_{-0.15}^{+0.17}$	  &5.3    &  $1.371/62$\\
       &  PC5 & 74463--76477	   & $0.64_{-0.19}^{+0.24}$  & $0.78_{-0.17}^{+0.18}$	  &1.3    &  $0.829/44$ \\
  \noalign{\smallskip}
  \hline
  \end{tabular}
  \end{center}
  \begin{list}{}{}
  \item[$^{\mathrm{a}}$]{Unabsorbed 2--10\,keV flux in units of $10^{-10}$ erg cm$^{-2}$ s$^{-1}$.}
  \item[$^{\mathrm{b}}$]{Cash statistics Cstat, and percentage of realizations ($10^{4}$ trials)
	with statistic $>$ Cstat. }
  \end{list}
  \end{table*}

Structured and complex peaks during outburst flares have already
been observed in other SFXTs [see e.g.\  
the flaring activity observed during the brightest phase of the outburst 
in IGR~J11215--5952 \citep{Romano2007}, or the several flares caught in 
XTE~J1739--302 with \sw\ \citep{Sidoli2008:sfxts_paperIII}, 
and with INTEGRAL at different epochs \citep{Blay2008}]. 
However it has never been observed spanning almost three orders 
of magnitude in flux, as found between the first two flare peaks in 2008 
(WT1 and PC3 data set in Fig.~\ref{igr08408:fig:2008xrttimes}, 
exceeding 10~counts\,s$^{-1}$) 
and their inter-flare emission (PC1, which goes down to $\sim0.01$ counts\,s$^{-1}$).
The hardness ratios (Fig.~\ref{igr08408:fig:2008lcv}d and
~\ref{igr08408:fig:2008xrttimes}b) 
show an anticorrelation with the source intensity, and our time-selected spectroscopy demonstrates 
that this evolution is produced by a change in the
absorbing column density, instead of a change in the source spectrum
(Fig.~\ref{igr08408:fig:2008xrtcontours}).
In particular, there is evidence for a progressively decreasing column
density with time (also during the low intensity emission of the part PC2 
of the light curve).
This is strongly suggestive of a ionization effect produced by the first
flare, resulting in a significant decrease in the measured column density
towards the source. 

During the first flare in 2008 BAT events were also available and a broad 
band spectrum could be extracted (see Table~\ref{igr08408:tab:timespecfits}); 
however, during the second and third flares BAT did not trigger 
(see Sect.~\ref{igr08408:spectra}) and the source was only detected at the 
$\sim 6\sigma$ level in survey data,  so no broad band coverage is available.
The broad band spectra during the two flares in 2006 and 2008 are
significantly different, especially as far as the absorbing column density
is concerned, which was about 20 times higher in the first flare in 2008 
than during the 2006 flare [where the absorption is consistent with the
interstellar value of 3$\times10^{21}$~cm$^{-2}$ derived from the optical
extinction to the optical counterpart HD~74194 
\citep{Leyder2007}]. 
We note that the 2006 flare spectrum was extracted about 2000~s after the
BAT trigger; \sw\ made a delayed slew to the source, too late to catch the 
peak of the first flare which triggered the BAT. 
This implies that the low column density observed
during the 2006 flare is probably so low when compared with the first 
2008 flare (WT1, taken only 137~s after the BAT trigger),
only because of the ionization effect caused by the flare that triggered the BAT 
(and not observed by XRT).
 
This suggests that, during extended bright flaring activity, it is not
possible to directly derive the mass of the clump
which is supposed to produce the enhanced accretion luminosity,
in the framework of the clumpy stellar wind model \citep{Walter2007},
because part or all of the accreting
material could have been ionized by a previous bright flare (as in 2006,
where the presence of a
precursor flare is deduced from the fact it triggered the BAT).

The broad-band
spectrum during the 2006 flare is well fitted with a power law with a 
high energy cutoff at $\sim11$\,keV.
This is compatible with a surface magnetic field of a typical X--ray
pulsar \citep{Coburn2002}, around 10$^{12}$~G,
much lower than that suggested by \citet{Bozzo2008}, or by
\citet{Gotz2007:08408-4503}.

The inter-flare emission is highly variable, showing a trend with an average 
decreasing flux after the first flare (for about 10~ks), then continuously increasing
by an order of magnitude (with a low intensity flaring activity, see PC2 part) up to the second flare.
The spectrum in this lower intensity inter-flare emission is well fitted with
a hard power law with a photon index of $\sim$1, which implies that accretion is still present.
The true quiescence has been observed by \citet{Leyder2007}, and displays a much 
softer spectrum and a lower luminosity around $\sim$10$^{32}$~erg~s$^{-1}$.
This inter-flare emission is probably due to the accretion of inter-clump medium, implying
a density (and/or a velocity) contrast in the supergiant wind of at least 10$^{3}$,
which is not unusual in line-driven winds.
Indeed,  in recent years evidence has been accumulated that hot
 stellar winds are not smooth, as previously thought,
but highly inhomogeneous  [see, e.g., 
\citet{Oskinova2007} and references therein], 
thus implying important consequences, for example, in the mass loss rate determination, 
and also in the stellar evolution.

	\subsection{Clumpy winds model: an estimate of the neutron
          star distance from the supergiant companion}

%
This new \sw\ data set interestingly allows to accurately determine the flare duration,
particularly of the second (PC3 data) and third flares (WT4/PC5).
Fitting their light curves with Gaussian profiles, we derive a FWHM of 370$\pm{20}$~s
and 500$\pm{30}$~s, respectively.
These durations can be used to derive information about the distance of the neutron
star from the supergiant companion during the outburst, since the clump sizes increase with the distance
from the supergiant star, as follows.
The size of a clump is determined by the balance pressure equation.
Following \citet{Lucy1980} and \citet{Howk2000},
the average density of a clump is:
\begin{equation} \label{density_clump_law}
\bar{\rho}_c = \rho_w(r) \left (  \frac{a_w^2 + C_{\rho}\omega^2}{a_c^2} \right)
\end{equation}
where $\rho_w(r)$ is the density profile of the homogeneous (inter-clump) wind, 
$a_w$ and $a_c$ are 
the inter-clump wind and the clump thermal velocity, respectively: $a_w^2 =  \frac{kT_w}{\mu m_H}$
and $a_c^2  =  \frac{kT_c}{\mu m_H}$, where $k$ is the Boltzmann constant, $T_w$ and $T_c$ are the
temperatures of the inter-clump wind and of the clump, respectively.
The constant $C_{\rho}=0.29$ accounts for the confining effect of the bow
shock produced by the ram pressure around the clump \citep{Lucy1980}, while  
$\omega$ is the relative velocity between the wind and the clump ($\omega = v_w - v_{c}$).

Wind structure simulations show that the average relative velocity
$\bar{\omega}$ is $\sim 5 \times 10^7$~cm~s$^{-1}$, the average
clump temperature is $\bar{T}_c \sim 10^5 \ K$, the average
temperature of the homogeneous inter-clump wind is $\bar{T}_w \sim 10^7 \ K$ 
\citep{Runacres2005}.
Adopting these values for $T_c$, $T_w$, $\omega$, and $\mu=1.3$ (for
solar abundances), we obtain: 
\begin{equation} \label{parte_costante}
\left (  \frac{a_w^2 + C_{\rho}\omega^2}{a_c^2} \right) \approx 200
\end{equation}

Since the density radial profile $\rho_w(r)$ of the homogeneous inter-clump wind is:
\begin{equation} \label{wind_density_law}
\rho_w(r) = \rho_w(R_{OB}) \frac{R_{OB}^2 v(R_{OB})}{r^2 v(r)}
\end{equation}
where $R_{OB}$ is the radius of the supergiant ($R_{OB}$=23.8~R$_{\odot}$, \citealt{Vacca1996}), then 
from equations (\ref{density_clump_law}), (\ref{parte_costante}), (\ref{wind_density_law})
we obtain:
\begin{equation} \label{clump_density_law_finale}
\bar{\rho}_{c}(r) = \bar{\rho}_{c}(R_{OB}) \frac{R_{OB}^2 v(R_{OB})}{r^2 v(r)}
\end{equation}
where $\bar{\rho}_{c}(R_{OB}) = \rho_w(R_{OB}) \times 200$.
Assuming a spherical geometry for the clumps, it is possible to obtain the expansion law of the
clump from equation
(\ref{clump_density_law_finale}): 
\begin{eqnarray}
R_{c}(r) = R_{c}(R_{OB}) \left ( \frac{r^2 v(r)}{R_{OB}^2 v_0} \right )^{1/3}
\end{eqnarray}
where $v_0 =v(R_{OB})$ is the initial velocity of the clump at the surface of the supergiant, which can 
be assumed $v_0$=10~km~s$^{-1}$ \citep{Castor1975}. For $v(r)$ we  consider the standard $\beta$ law for the 
velocity profile in line-driven winds, with a terminal velocity of 
1900~km~s$^{-1}$ and an exponent $\beta$=0.8.

The flare duration $t$ is linked with the thickness of the
clump, therefore 
\begin{equation} \label{equation_t_vs_distance}
t \approx \frac{2 R_{c}}{v_{c}} = \frac{2 R_{c}(R_{OB})}{v_{c}}
\left ( \frac{r^2 v(r)}{R_{OB}^2 v_0}  \right )^{1/3}
\end{equation}
where we assumed R$_{c}$(R$_{OB}$) $\approx$ 2$\times$10$^9$~cm; 
thus, flare durations of the order of $\sim$500~s can be produced at a radial distance 
of $\sim$$5 \ R_{OB}$ (or 10$^{13}$~cm), which implies an orbital period of $\sim$35~days, 
assuming a companion mass of 30~M$_\odot$.  This estimate is in agreement with that obtained
by \citet{Leyder2007}, but is significantly shorter than what calculated by \citet{Gotz2007:08408-4503}.

	\subsection{The orbital period and the outbursts double periodicity}
%
We note that the first three outbursts reported in literature (on 52821 MJD, 53870 MJD and 54012 MJD)
are indeed spaced by a multiple of 35 days (30 and 34 cycles, respectively, after the first observed outburst on 
52821 MJD).
On the other hand, the 2008 outburst (54652 MJD) does not occur 
strictly after an integer number of cycles with this assumed period. 
This suggested to try different combinations of two numbers (two sub-periods P$_{1}$ and P$_{2}$, such that
P=P$_{1}$+P$_{2}$=35 days), to test the possibility, suggested in \citet{Sidoli2007}, that the 
SFXTs outbursts occur with a double periodicity, when the neutron star crosses the equatorial wind of the companion
twice along the eccentric orbit (see their fig.~9c). In this hypothesis, the times
of the n$_{th}$ outburst is given by t$_n$=t$_0$+(n$_1$~P$_{1}$)+(n$_2$~P$_{2}$).
Among the several couples of numbers we tried (with sum equal to 35 days), we find 
that P$_{1}$=11~days and  P$_{2}$=24~days 
well account for all the four reported outburst times (calculated with  n$_1$=n$_2$=30,  n$_1$=n$_2$=34 and 
with  n$_1$=53 and n$_2$=52, respectively for the three outbursts, assuming t$_0$=52821 MJD), within
an uncertainty of 1 day.

	\subsection{\src\ in the SFXT context}

\begin{figure}
\begin{center}
\centerline{\includegraphics[width=9cm,angle=0]{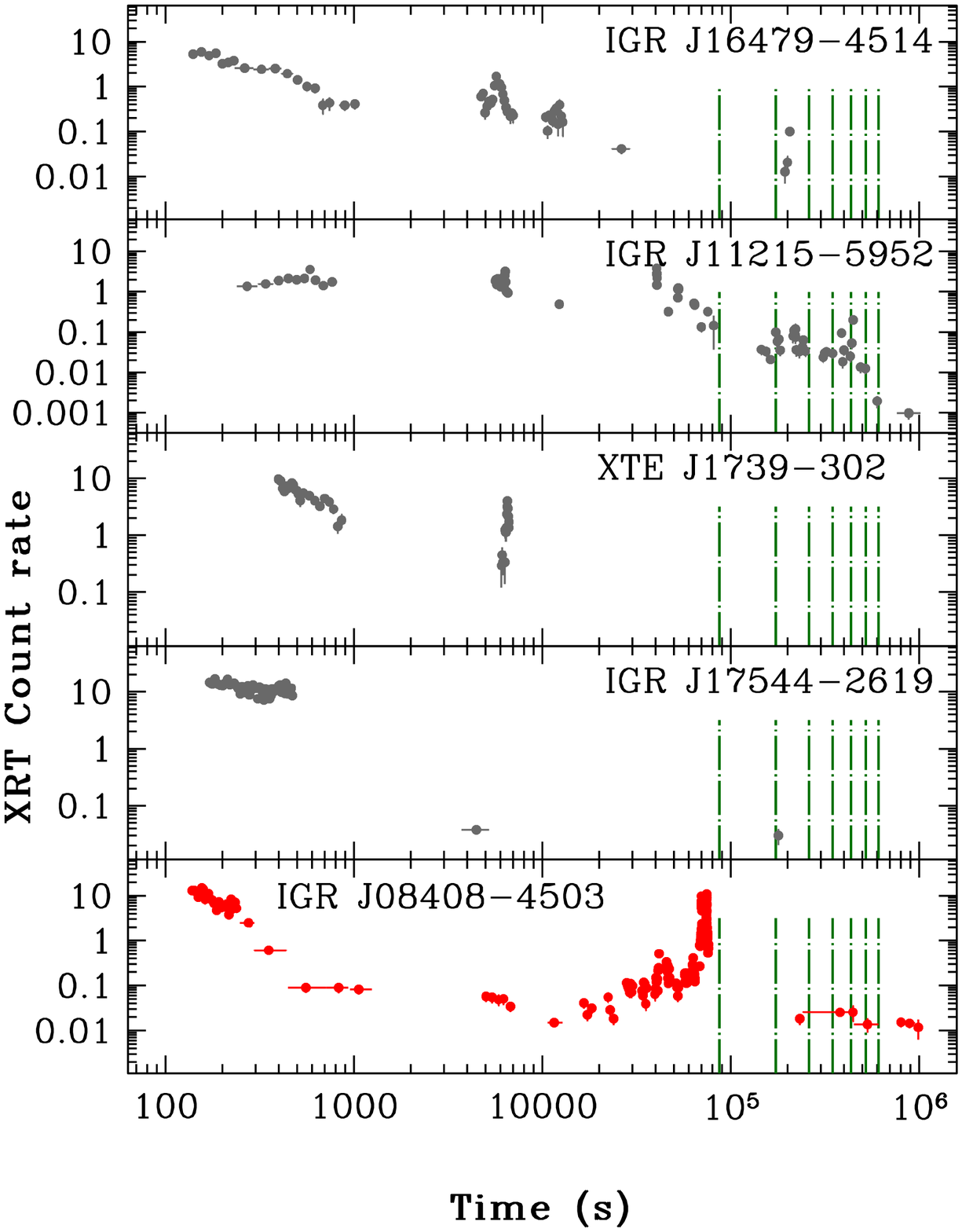}}
\caption{Light curves of the outbursts of SFXTs followed by {\it Swift}/XRT
referred to their respective triggers. We show the 2005 outburst of IGR~J16479$-$4514
\citep{Sidoli2008:sfxts_paperI}, which is more complete than the one observed in 2008 
\citep{Romano2008:sfxts_paperII}.
The IGR~J11215$-$5952 light curve has an arbitrary start time, since
the source
did not trigger the BAT (the observations were obtained as a ToO; \citealt{Romano2007}). 
The data on XTE~J1739$-$302 and IGR~J17544$-$2619 are presented in \citet{Sidoli2008:sfxts_paperIII}.
Note that where no data are plotted, no data were collected. 
For clarity, we drew dashed vertical lines to mark each day (up to one week) since the trigger. 
	}
\label{igr08408:fig:comp}
\end{center}
\end{figure}

The multiple flares we observe here, reaching similar peak luminosities, 
cover about one day of bright outburst emission
and are then followed by lower X--ray emission in the subsequent days,
similar to what we already observed in four other SFXTs with \sw\ (see
Fig.~\ref{igr08408:fig:comp}). 
The three flares observed here so close in time, after a long period
during which no bright flares have been observed, are suggestive of the 
fact that the neutron star is probably crossing
a region of higher average density of the wind, compatible with the model we
proposed for IGR~J11215$-$5952 \citep{Sidoli2007}. 
In this hypothesis, the outbursts in SFXTs are produced when the neutron star
crosses an equatorially enhanced region of the supergiant companion, which
is inclined with respect to the orbital plane. 
The different flares then are probably produced by inhomogeneities 
in the structure of the wind, which, in our proposed model, is anisotropic, 
showing a preferential plane along the equator of the
supergiant companion. 
This could explain why the flares happen in a preferential region
of the neutron star orbit, displaying a particularly high X--ray flare rate. 
New observations during the next outbursts are crucial in order to test
this hypothesis and  to confirm whether multiple 
outbursts are indeed a usual property of this (and other) SFXTs, and if there is
indeed an orbital phase dependence of the X--ray flare rate.

\section*{Acknowledgments}
We thank the \sw\ team for making these observations possible,
the duty scientists, and science planners. 
We also thank the anonymous referee for comments that helped improve the paper. 
This work was supported in Italy by contracts ASI/INAF I/088/06/0 and I/023/05/0. 
APB, PAE, KLP acknowledge STFC support. H.A.K.\ was supported by the \sw\ project. 
DNB and JAK acknowledge support from NASA contract NAS5-00136.

\bsp

\label{lastpage}


\begin{thebibliography}{}

\bibitem[\protect\citeauthoryear{{Bamba}, {Yokogawa}, {Ueno}, {Koyama} \&
  {Yamauchi}}{{Bamba} et~al.}{2001}]{Bamba2001}
{Bamba} A.,  {Yokogawa} J.,  {Ueno} M.,  {Koyama} K.,    {Yamauchi} S.,  2001,
  \pasj, 53, 1179

\bibitem[\protect\citeauthoryear{{Blay}, {Mart{\'{\i}}nez-N{\'u}{\~n}ez},
  {Negueruela}, {Pottschmidt}, {Smith}, {Torrej{\'o}n}, {Reig}, {Kretschmar} \&
  {Kreykenbohm}}{{Blay} et~al.}{2008}]{Blay2008}
{Blay} P.,  et al.,  2008, \aap, in press, ArXiv:0806.4097

\bibitem[\protect\citeauthoryear{{Bozzo}, {Falanga} \& {Stella}}{{Bozzo}
  et~al.}{2008}]{Bozzo2008}
{Bozzo} E.,  {Falanga} M.,    {Stella} L.,  2008, \apj, 683, 1031

\bibitem[\protect\citeauthoryear{{Burrows}, {Hill}, {Nousek}, {Kennea},
  {Wells}, {Osborne}, {Abbey} \& {Beardmore et al.}}{{Burrows}
  et~al.}{2005}]{Burrows2005:XRTmnras}
{Burrows} D.~N., et al.,   2005, Space Sci.\ Rev., 120, 165

\bibitem[\protect\citeauthoryear{{Cash}}{{Cash}}{1979}]{Cash1979}
{Cash} W.,  1979, \apj, 228, 939

\bibitem[\protect\citeauthoryear{{Castor}, {Abbott} \& {Klein}}{{Castor}
  et~al.}{1975}]{Castor1975}
{Castor} J.~I.,  {Abbott} D.~C.,    {Klein} R.~I.,  1975, \apj, 195, 157


\bibitem[\protect\citeauthoryear{{Coburn}, {Heindl}, {Rothschild}, {Gruber},
  {Kreykenbohm}, {Wilms}, {Kretschmar} \& {Staubert}}{{Coburn}
  et~al.}{2002}]{Coburn2002}
{Coburn} W.,  {Heindl} W.~A.,  {Rothschild} R.~E.,  {Gruber} D.~E.,
  {Kreykenbohm} I.,  {Wilms} J.,  {Kretschmar} P.,    {Staubert} R.,  2002,
  \apj, 580, 394

\bibitem[\protect\citeauthoryear{{Evans}, {Beardmore}, {Page}, {Tyler},
  {Osborne}, {Goad}, {O'Brien}, {Vetere}, {Racusin}, {Morris}, {Burrows},
  {Capalbi}, {Perri}, {Gehrels} \& {Romano}}{{Evans}
  et~al.}{2007}]{Evans2007:repository}
{Evans} P.~A.,  et al.,   2007, \aap, 469, 379

\bibitem[\protect\citeauthoryear{{Godet} et al}{2008}]{Godet2008:xrtcalib}
Godet O.,  et al.,   2008, \aap, submitted 

\bibitem[\protect\citeauthoryear{{G{\"o}tz}, {Falanga}, {Senziani}, {De Luca},
  {Schanne} \& {von Kienlin}}{{G{\"o}tz} et~al.}{2007}]{Gotz2007:08408-4503}
{G{\"o}tz} D.,  {Falanga} M.,  {Senziani} F.,  {De Luca} A.,  {Schanne} S.,
  {von Kienlin} A.,  2007, \apjl, 655, L101

\bibitem[\protect\citeauthoryear{{G{\"o}tz}, {Schanne}, {Rodriguez}, {Leyder}, {von
  Kienlin}, {Mowlavi} \& {Mereghetti}}{{G{\"o}tz}
  et~al.}{2006}]{Gotz2006:08408-4503discovery}
{G{\"o}tz} D.,  {Schanne} S.,  {Rodriguez} J.,  {Leyder} J.-C.,  {von Kienlin} A.,
  {Mowlavi} N.,    {Mereghetti} S.,  2006, \ATel, 813


\bibitem[\protect\citeauthoryear{{Howk}, {Cassinelli}, {Bjorkman} \&
  {Lamers}}{{Howk} et~al.}{2000}]{Howk2000}
{Howk} J.~C.,  {Cassinelli} J.~P.,  {Bjorkman} J.~E.,    {Lamers}
  H.~J.~G.~L.~M.,  2000, \apj, 534, 348



\bibitem[\protect\citeauthoryear{{in't Zand}}{{in't Zand}}{2005}]{zand2005}
{in't Zand} J.~J.~M.,  2005, \aap, 441, L1

\bibitem[\protect\citeauthoryear{{Kennea} \& {Campana}}{{Kennea} \&
  {Campana}}{2006}]{Kennea2006:08408-4503}
{Kennea} J.~A.,  {Campana} S.,  2006, \ATel, 818

\bibitem[\protect\citeauthoryear{{Leyder}, {Walter}, {Lazos}, {Masetti} \&
  {Produit}}{{Leyder} et~al.}{2007}]{Leyder2007}
{Leyder} J.-C.,  {Walter} R.,  {Lazos} M.,  {Masetti} N.,    {Produit} N.,
  2007, \aap, 465, L35

\bibitem[\protect\citeauthoryear{{Lucy} \& {White}}{{Lucy} \&
  {White}}{1980}]{Lucy1980}
{Lucy} L.~B.,  {White} R.~L.,  1980, \apj, 241, 300


\bibitem[\protect\citeauthoryear{{Markwardt}, {Barthelmy}, {Cummings},
  {Hullinger}, {Krimm} \& {Parsons}}{{Markwardt} et~al.}{2007}]{BATmanual2007}
{Markwardt} C.~B.,  {Barthelmy} S.,  {Cummings} J.~R.,  {Hullinger} D.,
  {Krimm} H.~A.,    {Parsons} A.,  2007, in ``The SWIFT BAT Software Guide
  (Version 6.3)'',
  http://swift.gsfc.nasa.gov/docs/swift/analysis/bat\_swguide\_v6\_3.pdf {The
  SWIFT BAT Software Guide (Version 6.3)}


\bibitem[\protect\citeauthoryear{{Masetti}, {Bassani}, {Bazzano}, {Dean},
  {Stephen} \& {Walter}}{{Masetti} et~al.}{2006}]{Masetti2006:08408-4503}
{Masetti} N.,  {Bassani} L.,  {Bazzano} A.,  {Dean} A.~J.,  {Stephen} J.~B.,
  {Walter} R.,  2006, \ATel, 815

\bibitem[\protect\citeauthoryear{{Mereghetti}, {Sidoli}, {Paizis} \&
  {Gotz}}{{Mereghetti} et~al.}{2006}]{Mereghetti:08408-4503}
{Mereghetti} S.,  {Sidoli} L.,  {Paizis} A.,    {Gotz} D.,  2006,
\ATel, 814

\bibitem[\protect\citeauthoryear{{Oskinova}, {Hamann} \&
  {Feldmeier}}{{Oskinova} et~al.}{2007}]{Oskinova2007}
{Oskinova} L.~M.,  {Hamann} W.-R.,    {Feldmeier} A.,  2007, \aap, 476, 1331

\bibitem[\protect\citeauthoryear{{Romano}, {Sidoli}, {Mangano}, {Mereghetti} \&
  {Cusumano}}{{Romano} et~al.}{2007}]{Romano2007}
{Romano} P.,  {Sidoli} L.,  {Mangano} V.,  {Mereghetti} S.,    {Cusumano} G.,
  2007, \aap, 469, L5

\bibitem[\protect\citeauthoryear{{Romano}, {Sidoli}, {Mangano}, {Vercellone},
  {Kennea}, {Cusumano}, {Krimm}, {Burrows} \& {Gehrels}}{{Romano}
  et~al.}{2008}]{Romano2008:sfxts_paperII}
{Romano} P.,  et al.,   2008,   \apjl, 680, L137

\bibitem[\protect\citeauthoryear{{Runacres} \& {Owocki}}{{Runacres} \&
  {Owocki}}{2005}]{Runacres2005}
{Runacres} M.~C.,  {Owocki} S.~P.,  2005, \aap, 429, 323

\bibitem[\protect\citeauthoryear{{Sguera}, {Barlow}, {Bird}, {Clark}, {Dean},
  {Hill}, {Moran}, {Shaw}, {Willis}, {Bazzano}, {Ubertini} \&
  {Malizia}}{{Sguera} et~al.}{2005}]{Sguera2005}
{Sguera} V.,  et al.,  2005, \aap, 444, 221

\bibitem[\protect\citeauthoryear{{Sidoli}, {Romano}, {Mangano}, {Cusumano},
  {Vercellone}, {Kennea}, {Paizis}, {Krimm}, {Burrows} \& {Gehrels}}{{Sidoli}
  et~al.}{2008a}]{Sidoli2008:sfxts_paperIII}
{Sidoli} L., et al.,  2008a, \apj, in press, ArXiv:0808.3085 

\bibitem[\protect\citeauthoryear{{Sidoli}, {Romano}, {Mangano}, {Pellizzoni},
  {Kennea}, {Cusumano}, {Vercellone}, {Paizis}, {Burrows} \&
  {Gehrels}}{{Sidoli} et~al.}{2008b}]{Sidoli2008:sfxts_paperI}
{Sidoli} L., et al.,   2008b, \apj, in press, ArXiv:0805.1808

\bibitem[\protect\citeauthoryear{{Sidoli}, {Romano}, {Mereghetti}, {Paizis},
  {Vercellone}, {Mangano} \& {G{\"o}tz}}{{Sidoli} et~al.}{2007}]{Sidoli2007}
{Sidoli} L.,  {Romano} P.,  {Mereghetti} S.,  {Paizis} A.,  {Vercellone} S.,
  {Mangano} V.,    {G{\"o}tz} D.,  2007, \aap, 476, 1307

\bibitem[\protect\citeauthoryear{{Smith}, {Heindl}, {Markwardt}, {Swank},
  {Negueruela}, {Harrison} \& {Huss}}{{Smith} et~al.}{2006}]{Smith2006aa}
{Smith} D.~M.,  {Heindl} W.~A.,  {Markwardt} C.~B.,  {Swank} J.~H.,
  {Negueruela} I.,  {Harrison} T.~E.,    {Huss} L.,  2006, \apj, 638, 974

\bibitem[\protect\citeauthoryear{{Swank}, {Smith} \& {Markwardt}}{{Swank}
  et~al.}{2007}]{Swank2007}
{Swank} J.~H.,  {Smith} D.~M.,    {Markwardt} C.~B.,  2007, \ATel, 999


\bibitem[\protect\citeauthoryear{{Vacca}, {Garmany} \& {Shull}}{{Vacca}
  et~al.}{1996}]{Vacca1996}
{Vacca} W.~D.,  {Garmany} C.~D.,    {Shull} J.~M.,  1996, \apj, 460, 914

\bibitem[\protect\citeauthoryear{{Vaughan} et~al.}{2006}]{vaughan2006:050315}
{Vaughan} S., {et~al.},  2006, \apj, 638, 920

\bibitem[\protect\citeauthoryear{{Walter} \& {Zurita Heras}}{{Walter} \&
  {Zurita Heras}}{2007}]{Walter2007}
{Walter} R.,  {Zurita Heras} J.,  2007, \aap, 476, 335

\bibitem[\protect\citeauthoryear{{Ward}, {Barthelmy}, {Beardmore}, {Brown},
  {Evans}, {Gehrels}, {Kennea}, {Krimm}, {Kuin}, {Markwardt}, {McLean},
  {Pagani}, {Page}, {Palmer}, {Perez}, {Romano}, {Sakamoto}, {Starling} \&
  {Tagliaferri}}{{Ward} et~al.}{2008}]{Ward2008:08408-4503}
{Ward} P.~A.,  et al.,  2008, \gcn, 7945, 1

\end{thebibliography}
\end{document}